\documentclass[aps,twocolumn,showpacs,nofootinbib]{revtex4}
\usepackage{mathrsfs}
\usepackage{pifont}
\usepackage{graphicx}
\usepackage{amsmath,amssymb,amsthm}
\usepackage[colorlinks=true,urlcolor=blue,linkcolor=blue,citecolor=red]{hyperref}
\newcommand{\beq}{\begin{equation}}
\newcommand{\eeq}{\end{equation}}
\newcommand{\bea}{\begin{eqnarray}}
\newcommand{\eea}{\end{eqnarray}}

\newcommand\lsim{\mathrel{\rlap{\lower4pt\hbox{\hskip1pt$\sim$}}
    \raise1pt\hbox{$<$}}}
\newcommand\gsim{\mathrel{\rlap{\lower4pt\hbox{\hskip1pt$\sim$}}
    \raise1pt\hbox{$>$}}}

\begin{document}

\title{ Metastable gauged O'Raifeartaigh}

\vspace{7mm}

\author{Borut Bajc$^{1}$ and 
Alejandra Melfo$^{1,2}$}
 
\address{$^{1}$ {\it\small J.\ Stefan Institute, 1000 Ljubljana, Slovenia}}
\address{$^{2}${\it\small Universidad de los Andes, 
Merida, Venezuela}}

\begin{abstract}
We study the possibility of obtaining metastable supersymmetry 
breaking vacua in a perturbative gauge theory without singlet fields, 
thus allowing for scenarios where a grand unified symmetry and  
supersymmetry are broken by the same sector. We show some explicit 
SU(5) examples. The minimal renormalizable example requires the use 
of two adjoints, but it is shown to inevitably lead to unwanted light 
states. We suggest various alternatives, and show that the viable 
possibilities consist of allowing for non-renormalizable operators, 
of employing four adjoints or of adding  at 
least one  field in a different representation.

\end{abstract}
 

\maketitle
 
\section{Introduction}
\label{section1}

The idea of trying to combine grand unified theories with supersymmetry 
breaking has been used already in the early days of supersymmetry 
\cite{Dine:1982zb,Banks:1982mg,Dimopoulos:1982gm,Kaplunovsky:1983yx} 
following mainly the suggestion of dimensional transmutation 
\cite{Witten:1981kv}. The tree order supersymmetry breaking vacuum 
enforced by the O'Raifeartaigh type superpotential automatically has 
a flat direction, which gets however stabilized at one loop exactly because 
of supersymmetry breaking corrections. All of these models are in a 
perturbative regime and make use of gauge singlets. The mediation 
of supersymmetry breaking to the MSSM sector is dominated by 
gravity, which cannot predict (although it can fit) the strong 
suppression of the flavour changing neutral currents. Later models 
\cite{Murayama:1997pb,Dimopoulos:1997ww, Luty:1997ny,
Agashe:1998kg, Agashe:2000ay} 
were able to get rid of gauge singlets, using nonperturbative gauge 
sectors to dynamically break supersymmetry. The minima here are not 
 global, but local and thus metastable, although with a long enough 
lifetime. A typical model has more sectors and gauge groups than 
usually assumed in phenomenological motivated models like MSSM 
or grand unification. The results are important and promising: the 
models considered are mainly realistic and quite natural (without 
fine-tunings), while the mediation is gauge dominated 
\cite{Dine:1981gu,Dine:1982zb, AlvarezGaume:1981wy, Dimopoulos:1982gm,
Dine:1994vc, Dine:1995ag, Giudice:1997ni, Giudice:1998bp}, 
an important result. 

What we want to explore in this work is the possibility to use as 
much as possible minimal gauge groups, no singlets and perturbative 
physics only. The best possibility (and the original motivation) is
to use a grand unified group $G$ (we will limit ourselves to SU(5)) 
without singlets and break both $G$ and $N=1$ supersymmetry spontaneously 
(an example of models which break $N=2$ supersymmetry spontaneously without 
the use of chiral singlets is given in \cite{Arai:2007md,Arai:2007ds}). 
At first glance this seems to be in contradiction with what we know 
from perturbative spontaneous supersymmetry breaking. In fact, 
 one needs a linear term in the superpotential, 
which must be a singlet, thus naively forbidding for it the use of 
a gauge multiplet.  
However, by choosing properly the basis it is easy 
to get rid of the linear term and 
thus have a form of the superpotential that can be directly employed in 
gauge theories without any need for singlets. This will be explicitly shown 
in section \ref{section2}. As it will be clear, such a construction 
is possible only because the considered vacuum is metastable 
(a recent revival of models with such vacua has been triggered by 
\cite{Intriligator:2006dd}). For such reasons we will call these models 
of the metastable gauged O'Raifeartaigh type. 
It is thus tempting to use this idea in realistic 
models like for example grand unified theories. 

Writing a superpotential that exhibits perturbative and spontaneous 
supersymmetry breaking without linear terms is only the first part 
of the story. The second part is to make these metastable gauged 
O'Raifeartaigh models realistic in the context of grand unified theories. 
The minimal SU(5) model will be explicitly presented in section 
\ref{section3}, together with the main virtues and drawbacks. 
The virtues are the fact that  two adjoint fields suffice  
to break both supersymmetry 
and SU(5) gauge symmetry spontaneously. We will show that the 
model is locally stable in some range of the vevs. One of the vevs is 
undetermined at tree order, and we will check that it can exhibit a 
metastable local minimum at one loop. The renormalizable superpotential 
has two terms only, a form which is enforced by a global U(1)$_R$ symmetry. 
The drawback of this simple example  is the presence of  light states, 
which makes it unrealistic. Possible corrections of this minimal 
scenario and the role of supergravity will be described in 
section \ref{section4}. 
We will present explicitly three realistic cases in which these 
unwanted light states are not present: 
1) the nonrenormalizable model with two ${\bf 24}$, section \ref{section4.1}, 
eq. (\ref{nonrenpot}); 
2) the renormalizable model with four ${\bf 24}$, section \ref{section4.2}, 
eq. (\ref{moreadjoints}); 
3) the renormalizable model with two ${\bf 24}$ and one ${\bf 75}$, 
section \ref{section4.3}, 
eq. (\ref{diffrepr}). Finally, some general 
remarks and a list of open problems (among which the suggestion to use this 
type of models in hybrid inflation without singlets) 
to be discussed in more detail 
elsewhere will be given in section \ref{section5}.

\section{From singlets to gauge multiplets}
\label{section2}

We start with the simplest model which exhibits metastable 
supersymmetry breaking following the general analysis 
\cite{Ray:2006wk}

\begin{equation}
W=S\left(\xi+\lambda\tilde\phi^2\right)\;.
\end{equation}

\noindent
It exhibits a tree level local minimum at 

\begin{equation}
\langle\tilde\phi\rangle=0\;\;\;,\;\;\;S\;{\rm undetermined}\;,
\end{equation}

\noindent
providing 

\begin{equation}
\left|\langle S\rangle\right|\ge\left|\frac{\xi}{2\lambda}\right|^{1/2}\;.
\end{equation}

\noindent
Such a superpotential cannot be directly written in terms of gauge 
multiplets, due to the existence of the linear term in $S$. 
It is however simple to get rid of it by redefining

\begin{equation}
\tilde\phi=\phi-\langle\phi\rangle\;,
\end{equation}

\noindent
and choosing $\langle\phi\rangle$ such that

\begin{equation}
\xi+\lambda\langle\phi\rangle^2=0\;.
\end{equation}

We end up with 

\begin{equation}
\label{muphis}
W=\mu\phi S+\lambda\phi^2S\;,
\end{equation}

\noindent
i.e., no linear terms, and with a local minimum at

\begin{equation}
\langle\phi\rangle=-\frac{\mu}{2\lambda}\;\;\;,\;\;\;S\;{\rm undetermined}\;,
\end{equation}

\noindent
provided it is in the allowed range 

\begin{equation}
\label{allowedsinglet}
\left|\langle S\rangle\right|\ge\frac{\left|\langle\phi\rangle\right|}
{\sqrt{2}}\;.
\end{equation}

This shows that one could start with eq. (\ref{muphis}), and since 
there are no linear terms in it, no singlet is really needed: 
both $S$ and $\phi$ in (\ref{muphis}) can be part of a gauge multiplet 
of a gauge group $G$, which vevs $\langle S\rangle$ and 
$\langle\phi\rangle$ break $G$ spontaneously to a subgroup $H$. 
In the next section we will 
give an SU(5) example with two adjoints, both breaking to 
SU(3)$\times$SU(2)$\times$U(1). 

\section{The simplest  example: two SU(5) adjoints }
\label{section3}

Using the results in the previous section, we can 
immediately write down a candidate 
for a metastable gauged O'Raifeartaigh SU(5) model:

\begin{equation}
\label{minimal}
W=\mu Tr\Sigma_1\Sigma_2+\lambda Tr\Sigma_1^2\Sigma_2\;.
\end{equation}

We expand the adjoints $\Sigma_i$ as
\begin{eqnarray}
\label{adjointdecomposition}
\Sigma_i=\begin{pmatrix}
  O_i+2\sigma_i/\sqrt{30}
& X_i
\cr
  \bar X_i
& T_i-3\sigma_i/\sqrt{30}
\cr \end{pmatrix}\;,
\end{eqnarray}
where  $\sigma_i$ are the Standard Model (SM) singlets,   $O_i$  the 
color octets  $(8,1; 0)$, $T_i$  the weak triplets $(1,3;0)$, and $X_i, 
\bar X_i $ the color triplet, weak doublets 
$(3,2; \pm5/3)$. 
The vev $v_1=\langle\sigma_1\rangle$ is obtained from

\begin{equation}
\left\langle\frac{\partial W}{\partial\sigma_1}\right\rangle=0\;\to\;
v_1=\frac{\sqrt{30}}{2}\frac{\mu}{\lambda}\;,
\end{equation}

\noindent
while supersymmetry breaking is signaled by a nonzero $F$ term:

\begin{equation}
\label{f2}
F_2^*\equiv\left\langle\frac{\partial W}
{\partial\sigma_2}\right\rangle=\frac{\lambda v_1^2}{\sqrt{30}}\;.
\end{equation}

The other vev, $v_2$ ($=\langle\sigma_2\rangle$), is 
undetermined at tree order, 
i.e. it is a flat direction. It will be stabilized 
by nontrivial 1-loop corrections to the K\" ahler potential, 
which at tree order is 

\begin{equation}
K_{0}=Tr\Sigma_i^\dagger\Sigma_i\;.
\end{equation}

\noindent
Since the vevs of the adjoints are diagonal, the D-terms are vanishing.

We have to check two things. 

First, that the above model does not 
contain tachyons. That the singlet has non-negative mass square 
at least for some choices of the vevs is expected from 
(\ref{allowedsinglet}). What remains to be checked are the 
masses of all other SM multiplets. One pair of the bosons  in $X_i, 
\bar X_i$  will 
provide the would-be Nambu-Goldstone bosons (mainly from $\Sigma_2$), 
while the other pair (mainly from $\Sigma_1$) will acquire a mass 
proportional to $v_2$, so we do not need to worry about them. 

After $SU(5)$ breaking, the singlet   in $\Sigma_1$ gets a supersymmetric mass

\beq
{\cal M}_{\sigma_1} = -\frac{2\lambda}{\sqrt{30}}v_2\:,
\eeq

\noindent
while the non-singlet mass matrices have in general the form

\begin{eqnarray} 
{\cal M}= \frac{\lambda}{\sqrt{30}} 
\begin{pmatrix}
c_2 v_2
& 
 c_1 v_1
\cr
 c_1 v_1
& 
  0
\cr
\end{pmatrix}\;,
\label{susymass}
\end{eqnarray}

\noindent
with $(c_1,c_2)=(6,4)$  for  color octets,  and  $(c_1,c_2)=(-4,-6)$  for weak 
triplets. 
The supersymmetry breaking mass terms in the Lagrangian are 

\begin{equation}
\delta L=\frac{\lambda F_2}{\sqrt{30}}
\left(- \sigma_1^2+ 2 O_1^2 - 3 T_1^2 - X_1 \bar X_1 \right)+h.c.
\end{equation}

One can now easily find out that there are no tachyonic states if 
the SM singlet scalar $\sigma_1$ is not tachyonic, which is true provided 
the analogue of (\ref{allowedsinglet}) is satisfied:

\begin{equation}
\label{constraint}
\left|v_2\right|\ge\frac{\left|v_1\right|}{\sqrt{2}}\;.
\end{equation}

The second thing we need to check is whether the flat direction $\sigma_2$ gets 
stabilized at 1-loop following the lines of \cite{Witten:1981kv}. 
All is needed is to check what happens with the wavefunction of the 
field that breaks supersymmetry ($\sigma_2$) \cite{Dimopoulos:1997ww}. 
In fact the potential at one loop gets corrected with respect to the tree order 
one by exactly the wavefunction renormalization (neglecting small finite 
corrections) through

\begin{equation}
\label{veff}
V(\sigma_2)\approx\frac{|F_2|^2}{Z_2(|\sigma_2|)}\;,
\end{equation}

\noindent
where $F_2$ can be read from (\ref{f2}) and $Z_2$ is the wavefunction 
renormalization at one loop. Obviously the minimum of the potential comes 
from the maximum of $Z_2$. At this point one can use the usual rules 
to write down the renormalization group equations - RGE's 
(a useful and concise set of rules can be found for 
example in \cite{Giudice:1997ni}). For the particle spectrum we take 
on top of the two adjoints just the minimal set of 
three generations of matter fields and one pair of $5_H$, $\overline{5}_H$ 
(the results can be easily generalized for more Higgs and/or messenger 
fields). We obtain ($\tau\equiv\frac{1}{8\pi^2}
\ln{\left(\frac{\mu}{M_{GUT}}\right)}$) the following system 

\begin{eqnarray}
\frac{d}{d\tau}g_5^{-2}&=&-2\;,\\
\frac{d}{d\tau}\ln{\lambda^2}&=&-30g_5^2+
21\lambda^2\;,\\
\label{z2}
\frac{d}{d\tau}\ln{Z_2}&=&10g_5^2-
\frac{21}{5}\lambda^2\;.
\end{eqnarray}

We have assumed that the couplings between the fundamental and 
adjoint Higgses are negligible \footnote{This assumption is consistent 
for example in the simplest 
of all cases, i.e. $W=\bar 5_H(y\Sigma_1+M)5_H$.}

The extremum of $Z_2$ fixes one parameter of the 
superpotential at the minimum

\begin{equation}
\label{lg5}
\lambda^2=\frac{50}{21}g_5^2\;.
\end{equation}

That the extremum of the potential is indeed a minimum can be 
seen from the negativity of the second derivative at the extremum

\begin{equation}
\label{d2z2dtau2}
\frac{1}{Z_2}\frac{d^2Z_2}{d\tau^2}=-180g_5^4\;.
\end{equation}

The minimum (and thus the GUT scale $v_2$) is determined by 
the equivalence (\ref{lg5}). 

We have thus checked that the Higgs sector (\ref{minimal}) can 
indeed break both SU(5) to the SM gauge group 
and supersymmetry. Also, the original parameters of the model 
($\mu$, $\lambda$) can be changed for the physical 
ones ($F$, $M_{GUT}$). 
Notice that all this has been achieved without any fine 
tuning of the model parameters. The gauge coupling was 
crucial in this game: the limit of gauge singlets would 
confirm the observation of \cite{Ray:2006wk} that 
metastable supersymmetry breaking vacua exist only when 
all values of the flat directions are allowed at tree order. In 
fact, for $g_5\to 0$ the one-loop correction would first push $v_2$ 
towards the origin, violating the bound (\ref{constraint}) 
and eventually finishing in one of the two supersymmetry 
preserving vacua $v_1=0$ or $v_1=\sqrt{30}\mu/\lambda$ 
(both with $v_2=0$).

The superpotential (\ref{minimal}) is the most general 
renormalizable superpotential for two SU(5) adjoints that 
satisfies a global U(1)$_R$ symmetry, under which $\Sigma_1$ 
is neutral and $\Sigma_2$ has charge $2$. This symmetry is 
spontaneously broken by the $v_2$ vev and has thus at the 
perturbative level an exact Nambu-Goldstone boson ($\sigma_2$). 
The R-symmetry must be eventually explicitly broken by 
supergravity corrections that cancel the cosmological 
constant \cite{Bagger:1994hh}, which will give a nonzero 
mass also to this pseudo-Nambu-Goldstone boson.

To summarize: SU(5) is broken at $v_2$, supersymmetry at $v_1$. 
The adjoint $\Sigma_1$ could in principle be used as a messenger.  

The model is simple and predictive, indeed too predictive, leading to 
inescapable problems. The most pressing one 
is that either the supersymmetry breaking scale is 
comparable to the GUT scale or there are light weak triplets and colour 
octets mainly from $\Sigma_2$. In fact from (\ref{susymass}) we can see 
that triplets and octets can have order $M_{GUT}$ 
mass only if $v_1={\cal O}(v_2)$, 
i.e. when $\sqrt{F}\approx v_1\approx v_2\approx M_{GUT}$. Since  
the most obvious candidate for the messengers are the MSSM 
multiplets in $\Sigma_1$, the typical soft mass is only loop 
(i.e. $\approx 10^{-2}$) 
suppressed with respect to the triplet and octet masses $\approx F/M_{GUT}$. 
Keeping $v_1$ as a free parameter one is still able to unify the gauge 
couplings, but at a too high scale slightly above $10^{19}$ GeV, with the 
sfermion and gaugino masses around $10^5$ GeV. Even if one 
accepted such a high scale, the calculation itself would turn out to be 
inconsistent, because $M_{GUT}\gsim 10^{19}$ GeV  would  make 
supergravity corrections to the soft masses dominant. Taking this into account 
consistently changes very little, making such a model unappealing. 
In the next section, we describe more realistic scenarios. 

\section{More realistic options}
\label{section4}

We see that the problem arises because the same scale that determines the 
light SM multiplets ($v_1$) specifies also the supersymmetry breaking 
$F\propto v_1^2$ 
and thus cannot be at the same time large and small. In order to provide for a 
different scale, one can resort basically to two possibilities: adding 
non-renormalizable interactions while keeping the field content minimal,  
or adding more fields and keep renormalizability.  We will find three different 
realistic models, described in sections \ref{section4.1}, \ref{section4.2} and 
\ref{section4.3} respectively. All three of them possess a global U(1)$_R$ 
symmetry, broken by the vacuum expectation value of the GUT field that gets 
a nonzero $F$ term. This is in accordance with the general theorem 
\cite{Nelson:1993nf}.

\subsection{Adding non-renormalizable operators}
\label{section4.1}

The first option is to keep the number of adjoints  at a minimum but increase 
the number of interaction terms, i.e. allow for non-renormalizable operators. 
Using higher powers in $\Sigma_1$ is still consistent with the U(1)$_R$ 
symmetry. The simplest correction 

\begin{equation}
W=Tr\left[\Sigma_2\left(\mu\Sigma_1+\lambda\Sigma_1^2+\frac{\alpha_1}{M}
\Sigma_1^3
+\frac{\alpha_2}{M}Tr\left(\Sigma_1^2\right)\Sigma_1\right)\right]
\label{nonrenpot}
\end{equation}

\noindent
is already enough: one can have large enough vev $v_1\approx v_2$ 
but with $F$ arbitrarily low (with a proper fine-tuning of the model 
parameters), as we now show.

From the equation of motion for $\sigma_1$, i.e. 
$\partial W/\partial\sigma_1=0$ 
we get

\begin{equation}
\mu=\frac{2\lambda}{\sqrt{30}}v_1-\frac{3}{M}\left(\frac{7}{30}\
\alpha_1+\alpha_2\right)v_1^2\;,
\end{equation}

\noindent
while the second equation $F^*=\partial W/\partial\sigma_2=0$ gives

\begin{equation}
\label{flambdac}
F^*=v_1^2\left[\frac{\lambda}{\sqrt{30}}-\frac{2}{M}
\left(\frac{7}{30}\alpha_1+\alpha_2\right)v_1\right]\;.
\end{equation}

This solution has no tachyonic states provided 

\begin{equation}
\label{v1bound}
2\left|\frac{v_2}{v_1}\right|^2\left|\frac{F^*}{v_1^2}-
\left(\frac{7}{30}\alpha_1+\alpha_2\right)\frac{v_1}{M}\right|\ge
\left|\frac{F^*}{v_1^2}\right|\;.
\end{equation}

For small enough $F$ this is always the case, so we do not need 
to worry anymore, allowing large values of $v_1$. Assuming all parameters 
real for simplicity we get for the determinants of the octet and triplet mass 
matrices

\begin{eqnarray}
\left(-\det{O}\right)^{1/2}&=&v_1\left[6\frac{F}{v_1^2}+
\left(\frac{75}{30}\alpha_1+10 \alpha_2\right)\frac{v_1}{M}\right]\;,\\
\left(-\det{T}\right)^{1/2}&=&v_1\left[4\frac{F}{v_1^2}+
\left(\frac{50}{30}\alpha_1+10 \alpha_2\right)
\frac{v_1}{M}\right]\;.
\end{eqnarray}

They depend very mildly on the supersymmetry breaking order parameter $F$. 
In the limit $F\to 0$ one has for $v_1\approx v_2\approx M_{GUT}$ 
the two eigenvalues of the order of $M_{GUT}^2/M$ (barring accidental 
cancellations). So, although there are intermediate states, they are much 
less harmful than the ones in the previous examples. 

There are various comments in order.

First, notice that in this example there is a fine-tuning needed to 
split the scales $v_1$ (that we want to be large in order to avoid 
too light states) and $\sqrt{F}$ (that we want to be small enough, 
possibly even around 100 TeV). This is seen 
for example from the constraint (\ref{flambdac}).

Another important point is that the cutoff $M$ cannot be too large 
for two reasons: first, eq. (\ref{flambdac}) tells us that $\lambda$ 
can be order $1$ as required by (\ref{lg5}) only for mild hierarchies 
$M_{GUT}/M$; second, one does not want too light intermediate states 
of mass $M_{GUT}^2/M$.

Finally, one could worry that the new operators introduced could influence 
the RGE's used to get the minimum of the effective potential. This is  not the 
case, since at one loop the $1/M$ suppressed operators do not contribute to 
the renormalization of the wave-functions.

Let us now check the gauge coupling unification constraints. The spectrum 
is the following: at $\Lambda_{SUSY}$ we have the MSSM superpartners 
(and the second Higgs), at $M_{GUT}^2/M$ we have two colour octets, two 
weak triplets and a pair of $X$, $\bar X$, i.e. 
one colour octet, one weak triplet and 
a full SU(5) adjoint. This is completely analogous to the case described in 
\cite{Bajc:2002pg} with the result that the final GUT scale is increased with 
respect to the usual MSSM case only by a factor 2, if we assume that the 
cutoff $M$ is 10 times the GUT scale. Due to the increase of $M_{GUT}$ and 
the appearence of an extra adjoint multiplet at the intermediate scale, the 
unification gauge coupling $\alpha_U$ increases by about $10\%$ with 
respect to the usual MSSM case.

\subsection{Adding more adjoints}
\label{section4.2}

If one wishes to stick to renormalizable models,  the simplest idea is to 
generalize  the model (\ref{minimal}) to something like

\begin{equation}
W=Tr\left[ \Sigma_{N+1}\left( \mu_i\Sigma_i+
\lambda_{ij}\Sigma_i\Sigma_j\right)\right]\;,
\end{equation}

\noindent
where now $i$ goes from $1$ to some integer $N$. Notice that 
$\lambda_{ij}$ in  general does not need to be symmetric, so in general 
a SU(5) invariant unitary rotation cannot 
diagonalize $\lambda$. For our purpose it is however enough 
to concentrate just on the diagonal elements of $\Sigma_{N+1}$ and $\Sigma_i$, 
so that these matrices commute and only the symmetric combination 
$\lambda_{ij}+\lambda_{ji}$ enters, which can be diagonalized. So 
we obtain in complete generality the $N$ replica of (\ref{minimal}), i.e. 

\begin{equation}
W=Tr\left[ \Sigma_{N+1}\left( \mu_i\Sigma_i+
\lambda_i\Sigma_i^2\right)\right]\;.
\end{equation}

Repeating the   exercise  in section \ref{section3}, we get 

\begin{eqnarray}
v_i=\frac{\sqrt{30}}{2}\frac{\mu_i}{\lambda_i}&;&
F_{N+1}=\sum_{i=1}^N\frac{\lambda_i}{\sqrt{30}}v_i^2\;.
\end{eqnarray}

In principle it could be possible to have large $v_i$ but small 
$F_{N+1}$ (by appropriate fine-tuning of the terms in the sum), but this 
cannot help, as we shall now see. The mass matrices that generalize 
(\ref{susymass}) are now $(N+1)\times (N+1)$ dimensional, and for the 
triplet and octet have the form

\begin{eqnarray}
\label{susymassgen}
{\cal M}= \frac{ 1}{\sqrt{30}} 
\begin{pmatrix}
\lambda_i c_{2,i}  \delta_{ij}v_{N+1}
& 
 \lambda_i c_{1,i }v_i
\cr
\lambda_i  c_{1,i} v_i
& 
  0
\cr
\end{pmatrix}\;,
\end{eqnarray}

\noindent
while the determinant is 


\begin{equation}
-\det{\left({\cal M}\right)}= 
\prod_{k=1}^N\left(\frac{\lambda_kc_{2,k}v_{N+1}}{\sqrt{30}}\right)
\sum_{i=1}^{N} 
\frac{\lambda_i c_{1,i}^2 v_i^2}{\sqrt{30}c_{2,i} v_{N+1}} \;.
\end{equation}

Since all the fields are adjoints, the Clebsch-Gordon coefficients 
$c_{1,i}, c_{2,i}$ are the same for  each SM  state,  and therefore 
the sum above is  proportional to $F_{N+1}$: in the limit 
$F_{N+1}\ll v_{N+1}^2$ we get $N$ masses of order 
$v_{N+1}$ and one of order $F_{N+1}/v_{N+1}$. Adding more adjoints in this 
way cannot give mass to the light colour octets and weak triplets.

This result is a consequence of the superpotential chosen, but there 
is at least 
another possibility. Namely, since any nonrenormalizable 
Lagrangian can be in principle obtained from a renormalizable 
one by integrating out heavy degrees of freedom, one could use directly the 
renormalizable potential that gives (\ref{nonrenpot}).  It turns out that, 
due to the 
linearity in $\Sigma_2$, not one but two additional adjoints ($\Omega_i$) are 
needed. The following ansatz 

\begin{eqnarray}
\label{moreadjoints}
W&=&-MTr\left(\Omega_1\Omega_2\right)+
Tr\left[\Omega_1\left(\mu_2\Sigma_2+\lambda_2\Sigma_2\Sigma_1\right)\right] 
\nonumber \\ & & +
Tr\left[\Omega_2\left(\mu_1\Sigma_1+\lambda_1\Sigma_1^2\right)\right]
\end{eqnarray}

\noindent
will do the job. One can show that this model has the right properties also 
in its renormalizable version (without integrating out $\Omega_{1,2}$) for 
 all the mass terms and couplings of order 1. Notice that there 
is still a U(1)$_R$ symmetry, under which $\Sigma_1$ and $\Omega_1$ have 
charge $0$ and $\Sigma_2$ and $\Omega_2$ have charge $2$. The model 
could presumably be generalized to 

\begin{equation}
W=Tr\left[\Sigma_2 \,f_\Sigma (\Sigma_1,\Omega_1)\right]+
Tr\left[\Omega_2 \,f_\Omega(\Sigma_1,\Omega_1)\right]\;.
\end{equation}

We will not push this model any further.

\subsection{Adding different representations}
\label{section4.3}

There is a further possibility to maintain renormalizability.
The point is that what precludes to have really 
different mass matrices of the MSSM adjoints and the singlet is the 
absence of enough terms in the superpotential. In other words, there 
is only one type of trilinear invariants for the adjoint fields (although 
for three different adjoints there are actually two such invariants, they 
are equivalent for diagonal elements that commute). So one can try 
to use different SU(5) representations, and the smallest one for this 
purpose to add to two adjoints is the ${\bf 75} $. One can write 
the most general renormalizable superpotential as

\begin{eqnarray}
\label{diffrepr}
W=&&\mu Tr(\Sigma_1 \Sigma_2) + \lambda_1 Tr(\Sigma_1^2 \Sigma_2) \nonumber\\
&+& \lambda_2 Tr(\Phi^2 \Sigma_2) +  \eta Tr(\Phi \Sigma_1 \Sigma_2)\;,
\end{eqnarray}

\noindent
where $\Phi$ is the  {\bf 75}. The supersymmetry breaking is achieved for  
the SM singlet vevs

\begin{eqnarray}
\langle \Phi \rangle &=& - \frac{5\sqrt{15}\eta}{16 \lambda_2} v_1 \;,\\
\langle \Sigma_1 \rangle \equiv v_1 &=& - \sqrt{\frac{15}{2}}\frac{\lambda_2 \mu }
{ \lambda_1\lambda_2 - \frac{125}{64} \eta^2} \;.
\end{eqnarray}

\noindent
We get now

\beq
F_2^* = - \frac{1}{\sqrt{30}\lambda_2}
\left(\lambda_1\lambda_2 - \frac{125}{64} \eta^2\right) v_1^2\;,
\eeq

\noindent
which can be fine-tuned to any desired value by fixing the expression 
in brackets. All one has to do now is to make sure that there are no light 
states, 
with masses proportional to the supersymmetry breaking parameter $F_2$. 

There are seven different states in all. Three of these are only present in 
{\bf 75}, namely the $(8,3;0)$, the $(3,1;\pm 10/3)$  and the 
$(6,2;\pm 5/3)$. It is evident from the superpotential that they get masses 
proportional to $\lambda_2 v_2$ since they do not mix. The $X,\bar X$ provide 
the Nambu-Goldstone bosons as before.  For the  other two, namely the color  
octets and weak triplets,  the 
determinant of the  supersymmetric mass matrices are

\begin{eqnarray}
 \det O  &=& - \frac{\sqrt{5} v_2 v_1^2}{\sqrt{6} \lambda_2 } \\
 &&\times
 \left(  \frac{16225}{18432} \eta^4 - 
 \frac{101 \sqrt{5}}{12\sqrt{6}} \eta^2 \lambda_2 \frac{F_2}{v_1^2} + 
 \frac{84}{5} \lambda_2^2 \frac{{F_2}^2}{v_1^4} 
 \right)   \nonumber \\
 \det T  &=&  v_1^2
 \left(  \frac{15 \sqrt{30}}{32}\frac{\eta^2}{\lambda_2} - 4 
 \frac{F_2}{v_1^2} \right)^2
 \end{eqnarray}

As can be seen, there are no light states left. Thus, this can be considered 
the minimal renormalizable version.

\subsection{Supergravity corrections}

In supergravity it is possible to spontaneously break supersymmetry and 
SU(5) with just one adjoint \cite{Bajc:2006pa}, although with considerable 
fine-tuning. In this paper we want to take the opposite limit, i.e. to avoid the 
domination of terms suppressed by the Planck mass. However, supergravity 
is there, if nothing else,  to cancel the cosmological constant. 
Here we will shortly check what supergravity does to our models. 
We will limit ourselves to the most delicate aspects of the above scenario, i.e. 
the stability of the minimum found through the RGE's and to the R-axion mass.

Consider the nonrenormalizable model with two adjoints. Although the 
model has a cutoff lower than the Planck scale, we assume that the 
UV completion at this cutoff, valid all the way to $M_{Pl}$, maintains 
at least approximately the form of the SM singlets' superpotential

\begin{equation}
W=F(\phi_i)\sigma_2+W_0(\phi_i)\;,
\end{equation}

\noindent
where $F(\langle\phi_i\rangle)\equiv F$ sets the scale of supersymmetry 
breaking and $W_0(\langle\phi_i\rangle)\equiv W_0\approx FM_{Pl}$ 
fine-tunes the cosmological constant to zero. Assuming that all vevs are 
smaller than $M_{Pl}$ the typical supergravity contribution to the 
potential for $\sigma_2$ is schematically $F^2(\sigma_2/M_{Pl})^n$ 
and so only the lowest $n$'s are relevant. The correction to the 
mass is

\begin{equation}
\label{sugra}
\Delta m_{\sigma_2}^2\approx\frac{4}{3} \left(\frac{F}{M_{Pl}}\right)^2\;,
\end{equation}

\noindent
to be compared with the mass found in the global supersymmetric case. 
This can be easily read off from (\ref{veff}) and (\ref{d2z2dtau2})

\begin{equation}
\label{susy}
m_{\sigma_2}^2\approx 360\left(\frac{\alpha_U}{4\pi}\right)^2
\left(\frac{F}{M_{GUT}}\right)^2\;.
\end{equation}

We see that the mass square from the solution (\ref{susy}) in the global 
supersymmetry case is numerically (for $M_{GUT}\approx 4.10^{16}$ GeV, 
$M_{Pl}\approx 2.10^{18}$ GeV, $\alpha_U\approx 1/20$) 
15 times or so bigger than the 
supergravity contribution (\ref{sugra}). So, the mass is stable. 

There is however a new linear contribution and this represents the main 
danger. The main part of the potential can be written schematically as 
the sum of the leading gauge contribution and the supergravity corrections

\begin{equation}
V\approx m_{\sigma_2}^2(\sigma_2-M_{GUT}^0)^2+\frac{F^2}{M_{Pl}}\sigma_2+...
\end{equation}

In the limit $M_{Pl}\to\infty$ we had 
$M_{GUT}^0=\langle\sigma_2\rangle\equiv M_{GUT}$, 
but now the true minimum gets 
shifted as (we omit numbers of order one)

\begin{equation}
\label{mgut}
M_{GUT}=M_{GUT}^0+\frac{F^2}{m_{\sigma_2}^2M_{Pl}}\;.
\end{equation}

The two contributions are of the same order and the supergravity one could 
even dominate. To settle it one would need to perform a 
more precise calculation. One can 
however notice that the value $M_{GUT}^0$ was defined as the scale, 
at which the equality (\ref{lg5}) is satisfied. But then it is enough to shift 
this scale to a different value, so that the final $M_{GUT}$ (\ref{mgut}) is 
what we would like it to be. 

Another issue is the R-axion mass. The constraint of a vanishing 
cosmological constant requires a constant term of order $FM_{Pl}$ 
in the superpotential. This term explicitly breaks the U(1) R-symmetry. 
The pseudo R-axion gets thus a non-vanishing mass of order 
\cite{Bagger:1994hh}

\begin{equation}
m_a^2\approx\frac{F^2}{M_{GUT}M_{Pl}}\;.
\end{equation}

Whether the model is cosmologically safe or not depends on the value 
of $F$. A weak scale R-axion mass is dangerous, for similar reasons as 
moduli, see for example \cite{Randall:1994fr, Dvali:1995mj,Dine:1995uk}
for possible solutions in this case. In the opposite case of small $F$ the 
R-axion mass is harmless.

A short comment is due on D-terms. It is known, that in general $N=1$ $D=4$ 
supergravity, the $D$ and $F$ terms are connected 
\cite{Binetruy:2004hh,Choi:2005ge}. This means, that if $D$-terms are 
non-zero, they are related to $F$-terms. In our case the adjoints 
are diagonal, so their $D$-terms are still zero, similar to the global limit.

\section{Conclusions}
\label{section5}

We have shown that it is possible to construct realistic superpotentials 
that break perturbatively both supersymmetry and a gauge symmetry without using 
singlets. This is possible only because the minima considered were 
metastable. For such models there is no reason to introduce extra 
gauge sectors, which dynamically break supersymmetry. One can thus 
study just simple gauge groups, a particularly appealing situation in 
case of grand unified theories. The price to pay is that extra states need 
to be introduced. 

We found three different realistic SU(5) examples: 

\begin{enumerate}
\item  nonrenormalizable model with two ${\bf 24}$, section \ref{section4.1}, 
eq. (\ref{nonrenpot}); 
\item  renormalizable model with four ${\bf 24}$, section \ref{section4.2}, 
eq. (\ref{moreadjoints}); 
\item   renormalizable model with two ${\bf 24}$ and one ${\bf 75}$, 
section \ref{section4.3}, eq. (\ref{diffrepr}).
\end{enumerate}

All three of them have a spontaneously broken U(1)$_R$ 
global symmetry.

There are many issues not touched in this paper, to be addressed in subsequent 
work. Let us mention some of them.

{\it The doublet-triplet problem.} In the minimal  case of 
the renormalizable model with two adjoints the doublets and triplets of a 
single pair of $5_H$ and $\bar 5_H$ cannot be split enough even with 
fine tuning. In fact, $\Sigma_2$ should not couple to the fundamentals, 
because its $F$ term destabilizes the weak scale. On the other side $\Sigma_1$ 
has a too small vev (of the order of $\sqrt{F}$) to split 
enough the doublets and triplets. 
It is thus reassuring that in the realistic versions this problem
disappears, since now $v_1$ can be of the order of the GUT scale. 

In models with $75$ one can use the missing partner mechanism 
\cite{Masiero:1982fe}. Such a model has quite some number of huge 
representations (two $24_H$, one $75_H$ and one pair of $50_H$ 
and $\overline{50}_H$), but it should be stressed that no fine-tuning is 
needed, except the obvious one that creates the hierarchy 
$\sqrt{F}\ll M_{GUT}$, needed in all known perturbative supersymmetry breaking 
models without light states. 

{\it Mediation of supersymmetry breaking.} The obvious mediators in  all 
these type of models are the heavy gauge bosons and the adjoints. They 
can dominate over gravity only for relatively low $M_{GUT}$, not much higher 
than the usual in MSSM. The large number of fields can help for this purpose. 
Notice that the potential problem of negative soft mass squared is not 
necessarily there due to the subsequent running, as shown recently in 
\cite{Dermisek:2006qj}. Other possible contributions need the introduction 
of extra (possibly intermediate scale) states, like the usual extra pairs 
of SU(5) fundamental and anti-fundamentals, or a pair of $15_H$ and 
$\overline{15}_H$ that can be used also for the neutrino 
masses \cite{Joaquim:2006uz}. 

{\it Non-perturbative contributions.} We have assumed that 
the perturbative part of the superpotential 
dominates. One could ask, how can the non-perturbative contributions 
influence the picture. Can one calculate them? 
The models considered are realistic and thus necessarily complicated 
enough to make the usual techniques (use of holomorphicity, symmetries, etc)
hard and probably non conclusive. Notice that none of the models we presented 
is ultraviolet free. The best one can do without further 
work is to make the most sensitive part of our 
mechanism, i.e. the presence of a U(1)$_R$ symmetry, independent on 
the quantum non-perturbative corrections. This can be guaranteed by 
making the U(1)$_R$ global symmetry non-anomalous. Of course this 
depends on the model chosen. For example, in the non-renormalizable 
model with two adjoints, one needs to add to the usual spectrum (two 
adjoint Higgses, a pair of fundamental Higgses and three generations 
of $10_F$ and $\bar 5_F$ matter) also two pairs of ($5_i$+$\bar 5_i$) 
chiral multiplets with vanishing $R$-charge (enforced for example by a term 
$\lambda_i\bar 5_i\Sigma_2 5_i$ in the superpotential). A general 
treatment of this issue is very interesting, but beyond the scope 
of this paper.

{\it Vacuum metastability.} We have assumed throughout the paper 
that the vacuum lifetime is longer 
than the age of the universe. This can be checked either with an explicit 
calculation using the full 1-loop effective potential, or estimated 
as it is done in \cite{Dimopoulos:1997ww}. 
Using the constraint (\ref{v1bound})  
and the method in \cite{Dimopoulos:1997ww} one finds for such a bounce 
action an approximate value of $S_B\approx 2\pi^2M_{GUT}^2/|F|$, which 
is much larger than the required value of $\approx 500$ needed for the 
lifetime to be longer than the age of the universe.

{\it Different gauge groups.} We have limited ourselves to the prototype 
example of a SU(5) grand unified theory. For many aspects the SO(10) 
GUT is more successful. Unfortunately the minimal renormalizable 
version \cite{Aulakh:2003kg} cannot break at the same time the gauge 
group and supersymmetry, the reason being the absence of a flat direction. 
An extra problem in such nonminimal groups is the need for breaking rank, 
which typically needs an extra fine-tuning. 

A special role can be played here by partial unified groups, like the 
Pati-Salam or the Left-Right group. Being possible at lower scales 
without being necessarily worried about proton decay constraints, 
they can automatically give a low enough supersymmetry breaking scale 
without any fine-tuning.  The minimal model with two fields  can work 
in both cases, however,  again an additional sector (and additional 
fine-tuning) is needed in order to break rank. Of course the whole 
motivation for supersymmetry is here less pronounced: no hierarchy problem 
because of  little or no  hierarchy, no one-step unification because 
of intermediate scales. 

{\it Inflation without singlets.} It is interesting that these type 
of models give possible candidates for a non-singlet (although still 
MSSM singlet) inflaton. Apart from few exceptions (for example 
\cite{Allahverdi:2006iq} in MSSM and \cite{Lazarides:1993fi} in a GUT) 
this would be one of the very few examples of such inflatons on the 
market. The simplest model (\ref{minimal}) is very similar to the 
prototype model of F-term hybrid inflation \cite{Dvali:1994ms}. If 
one is not too ambitious and does not pretend that the same model 
describes also supersymmetry breaking, this simple model could in 
principle work. In fact, in order to get rid of the unwanted light 
states, one can think that the final state after inflation is in the 
true minimum, in which both adjoints become heavy. 
Preliminary results seem to confirm that inflation can indeed take place, 
in  a similar manner as in  the case with singlets introduced in 
\cite{Dvali:1994ms}. For example, 
one can calculate the derivatives of the 1-loop potential and find out 
that the usual requirements for inflation to happen are satisfied. 
What would be particularly interesting is to see if there are any 
differences in predictions with respect to the case with a singlet. 
This work is in progress and a detailed analysis will be presented elsewhere.

\section*{Acknowledgements}

We thank Charanjit S. Aulakh, Lotfi Boubekeur, Ilja Dor\v sner, Gia Dvali, 
Riccardo Rattazzi and Goran Senjanovi\' c for discussions. 
This work has been supported by the Slovenian Research Agency.

\end{document}